\documentclass[twocolumn,aps,pre,floatfix,nofootinbib]{revtex4-2}
\usepackage{psfrag,epsfig,amsfonts,amssymb,amsmath,wasysym,bm}
\usepackage{dcolumn}
\usepackage{bbold}
\usepackage{color}
\usepackage{soul}
\usepackage[normalem]{ulem}
\usepackage{tabularx}
\usepackage{tikz}

\usepackage[inline]{enumitem} 

\usepackage{hyperref}

\newcommand{\kB}{k_{\mathrm B}} 

\newcommand{\RR}{{\mathbb R}}
\newcommand{\CC}{{\mathbb C}}
\newcommand{\NN}{{\mathbb N}}

\newcommand{\tr}{\mbox{Tr}}
\newcommand{\hr}{{\cal H}}

\providecommand{\norm}[1]{\|#1\|}
\providecommand{\opnorm}[1]{\|#1\|_{\!\!\; op}}

\newcommand{\lmat}{\left( \begin{matrix}}	
\newcommand{\rmat}{\end{matrix} \right)}	

\begin{document}

\title{Onsager's regression hypothesis adjusted to quantum systems}

\author{Peter Reimann}
\author{Christian Eidecker-Dunkel}
\affiliation{Faculty of Physics, 
Bielefeld University, 
33615 Bielefeld, Germany}
\date{\today}

\begin{abstract}
Onsager's regression hypothesis 
connects 
the 
temporal relaxation 
of close-to-equilibrium systems 
with
their dynamical correlation functions at thermal 
equilibrium.
While the hypothesis is provably correct in classical systems, 
it is known to fail in the quantum regime.
Here, we derive a suitably adjusted quantum 
version of Onsager's original hypothesis.
Rigorous analytical results are complemented by a variety of 
numerical examples.
\end{abstract}

\maketitle

\section{Introduction and main results}
\label{s1}
Onsager originally postulated and utilized the so-called 
regression hypothesis to establish his celebrated reciprocal 
relations between the kinetic coefficients of
irreversible processes close to thermal equilibrium
\cite{ons31}.
Qualitatively,
the basic physical content of the hypothesis is
that the 
temporal correlations
of a system at thermal equilibrium 
are sufficient to describe how a slightly 
perturbed system returns to equilibrium.
The quantitative formulation of the hypothesis
will be provided later (see Eq. (\ref{15}) below).
For classical systems,
Onsager's hypothesis has been rigorously
deduced from 
a microscopic description
for 
instance in Ref. \cite{tak52}.
In the realm of quantum mechanics,
the hypothesis is known to fail, as 
pointed out for example in Refs. \cite{tal86,for96,shi17}
(see also Sec. \ref{s5} below).
The main objective of our present work is to deduce a
properly modified quantum version of Onsager's regression hypothesis 
from basic microscopic considerations
(see Eq. (\ref{18}) below).

Concretely, let us
consider a quantum 
system with Hamiltonian $H$ 
at thermal equilibrium, described by a canonical ensemble
\begin{eqnarray}
\rho & := & Z^{-1}\, e^{-\beta H}
\, ,
\label{1}
\\
Z & := & \tr\{e^{-\beta H}\}
\, ,
\label{2}
\\
\beta & := & 1/\kB T
\ ,
\label{3}
\end{eqnarray}
where $T$ is the system's temperature and $\kB$ Boltzmann's constant.
The corresponding thermal equilibrium expectation 
value of an observable (Hermitian operator) $A$ is denoted as
\begin{eqnarray}
A_{th} :=\tr\{\rho A\}
\ .
\label{4}
\end{eqnarray}
Similarly, 
the temporal correlation
(also called, among others, dynamic or 
two-point correlation function)
of two Hermitian operators $V$ and $A$ at thermal equilibrium 
is given by
\begin{eqnarray}
C_{\! V\!\!A}(t):=\tr\{\rho\;\! V\! A(t)\} - V_{th} A_{th}
\ ,
\label{5}
\end{eqnarray}
where $A(t)$ is the observable $A$ at time $t$ in the
Heisenberg picture,
\begin{eqnarray}
A(t)
& := &
e^{iH t/\hbar} A e^{-iHt/\hbar}
\ .
\label{6}
\end{eqnarray}
In general, $C_{\! V\!\!A}(t)$ in (\ref{5})
is a complex valued function,
whose real and imaginary parts are denoted as
\begin{eqnarray}
R_{\! V\!\!A}(t)
& := &
\mbox{Re}(C_{\! V\!\!A}(t))
\ ,
\label{7}
\\
I_{\! V\!\!A}(t)
& := &
\mbox{Im}(C_{\! V\!\!A}(t))
\ .
\label{8}
\end{eqnarray}

Next we turn to a slightly different situation.
Namely, the system is prepared at time $t=0$
in an initial state of the form
\begin{eqnarray}
\rho_0 & := & Z_{\! g}^{-1}\, e^{-\beta H_{\! g}}
\, ,
\label{9}
\\
Z_{\! g} & := & \tr\{e^{-\beta H_{\! g}}\}
\label{10}
\end{eqnarray}
corresponding to the canonical ensemble 
of a modified Hamiltonian
\begin{eqnarray}
H_{\! g}:=H - g V
\ ,
\label{11}
\end{eqnarray}
where $V$ is a perturbation operator and 
$g$ a small parameter.
While $\rho_0$ in (\ref{9}) 
amounts to a thermal equilibrium state with respect to
the modified Hamiltonian $H_{\! g}$, it is a non-equilibrium 
state with respect to the actual Hamiltonian $H$ 
of the system we are considering.
Accordingly, this initial state then evolves for $t>0$ 
into the state 
\begin{eqnarray}
\rho(t) = e^{-iHt/\hbar}\rho_0\,e^{iHt/\hbar}
\ ,
\label{12}
\end{eqnarray}
yielding for an observable 
$A$ the time-dependent expectation values
\begin{eqnarray}
\langle A\rangle_t :=\tr\{\rho(t) A\} \, .
\label{13}
\end{eqnarray}
Equivalently, 
they can be rewritten in the Heisenberg picture 
by means of (\ref{6}) and (\ref{12}) as
\begin{eqnarray}
\langle A\rangle_t & = & \tr\{\rho_0 A(t)\} 
\, ,
\label{14}
\end{eqnarray}

In terms of the so-defined quantities, Onsager's regression 
hypothesis assumes the form
\begin{eqnarray}
\langle A\rangle_t - A_{th}= g \beta\, R_{\! V\!\!A}(t)
\label{15}
\end{eqnarray}
for 
sufficiently weak perturbations, i.e., 
up to corrections of order $g^2$.
As announced, it 
relates the temporal correlations at thermal equilibrium
to the time-dependent expectation values
of a system close to equilibrium
(usually exhibiting some kind of relaxation).

For quantum systems as we consider them here,
Onsager's regression hypothesis
(\ref{15}) is known to be incorrect
\cite{tal86,for96,shi17}
(a simple analytical example will be provided in Sec. \ref{s5}).
However, an appropriately adjusted version of such a 
relation has to our knowledge
never been proposed until now.
This is the main objective of our present work.
Namely, we will derive for sufficiently small $g$
the following modified relation:
\begin{eqnarray}
 \langle A\rangle_t - A_{th}
 & = &  
g \beta \sum_{k=0}^\infty  \frac{(i\tau_B)^k}{(k+1)!}\,  C^{(k)}_{V\!A}(t)
\ ,
\label{16}
\end{eqnarray}
where $f^{(k)}(t)$ indicates the $k$-th derivative of 
any given function $f(t)$,
and where
\begin{eqnarray}
\tau_B:=\hbar \beta
\label{17}
\end{eqnarray}
is the so-called Boltzmann time. 
Furthermore, we will show that the sum on the right
hand side of (\ref{16}) is a purely real function of $t$
(even though 
the single summands may possibly be complex).
Exploiting the definitions in (\ref{7}) and (\ref{8})
one thus
can conclude that
\begin{eqnarray}
 \langle A\rangle_t -
 A_{th} =
g \beta \sum_{k=0}^\infty \frac{\tau_B^k}{(k+1)!} \,S_k(t)
\, , \ \ \ 
\label{18}
\end{eqnarray}
where
\begin{eqnarray}
S_k(t) :=  
\left\{
  \begin{array}{l@{\hspace{3ex}}ll}
    (-1)^{k/2}\,R^{(k)}_{\! V\!A}(t)  & \mbox{for even $k$,}\\[0.1cm]
    (-1)^{\frac{k+1}{2}} \,I^{(k)}_{\! V\!A}(t)  & \mbox{for odd $k$.}\\
  \end{array}
\right.
\label{19}
\end{eqnarray}
In particular, Onsager's hypothesis (\ref{15}) is asymptotically 
recovered for small values of $\tau_B$, 
commonly considered as corresponding to the classical limit
in view of (\ref{17}).

The rest of the paper is organized as follows.
The derivation of our main results (\ref{16})-(\ref{19})
is provided in Sec. \ref{s2}.
The validity range of these results
is established in Sec. \ref{s3},
complemented by the more 
rigorous details in Appendix \ref{app1}.
Some general implications of physical interest
are discussed in Sec. \ref{s4},
followed by various numerical examples in  Sec. \ref{s5},
{a modification/extension of our main results 
in Sec. \ref{s6} and Appendix \ref{app2},}
and some concluding remarks in Sec.\ref{s7}.

\section{Derivation of the main results}
\label{s2}
In essence, the derivation of our main results
(\ref{16})-(\ref{19}) is relatively easy.
Additional details are provided in Sec.~\ref{s3} and Appendix \ref{app1}.

Employing the definitions
\begin{eqnarray}
\psi(\lambda)
&:=&
e^{\lambda H} e^{-\lambda H_{\! g}}
\ ,
\label{20}
\\
q & := & \tr\{\rho\, \psi(\beta)\}
\ ,
\label{21}
\end{eqnarray}
it follows with (\ref{1}) and (\ref{9}) that
\begin{eqnarray}
q = Z_{\! g}/Z
\ ,
\label{22}
\end{eqnarray}
and with (\ref{14}) that
\begin{eqnarray}
\langle A\rangle_t 
=
q^{-1}\, \tr\{\rho \, \psi(\beta) A(t)\}
\, .
\label{23}
\end{eqnarray}
Furthermore, 
we can conclude from (\ref{20}) and (\ref{11})
that
\begin{eqnarray}
\psi'(\lambda) 
& := & d \psi(\lambda)/d\lambda= 
e^{\lambda H} H e^{-\lambda H_{\! g}} - e^{\lambda H} H_{\! g} e^{-\lambda H_{\! g}} 
\nonumber
\\
& = & e^{\lambda H} g V e^{-\lambda H_{\! g}}
= g V_\lambda \psi(\lambda)
\, ,
\label{24}
\\
V_\lambda & := & e^{\lambda H} V e^{-\lambda H}
\, .
\label{25}
\end{eqnarray}
Integrating (\ref{24}) and exploiting that $\psi(0)=1$ yields
\begin{eqnarray}
\psi(\lambda) = 1 + g \int_0^\lambda dx \, V_x \, \psi(x)
\ .
\label{26}
\end{eqnarray}
Upon iteration, we thus obtain
\begin{eqnarray}
\psi(\lambda) = 
1 
+g  \!\int\limits_0^\lambda \! dx \, V_x
+  g^2  \!\int\limits_0^\lambda\! dx \, V_x \!\int\limits_0^x \! dy \, V_y \psi(y) \, .
\ \ \ 
\label{27}
\end{eqnarray}
Together with (\ref{1}), (\ref{21}), and (\ref{25}) this implies
\begin{eqnarray}
q & = &1+g\,\beta\, \tr\{\rho V\} + g^2 Q
\ ,
\label{28}
\\
Q & := &  \!\int_0^\beta\! dx \,  \!\int_0^x \! dy \, \tr\{\rho V_x V_y \psi(y) \}
\ .
\label{29}
\end{eqnarray}
Likewise, one finds that
\begin{eqnarray}
\tr\{\rho \, \psi(\beta) A(t)\}
=
\tr\{\rho A(t)\} + g\! \int\limits_0^\beta d\lambda\,  Y_t(\lambda) + g^2 R_t
\,, \ \ \ \ \ \ 
\label{30}
\end{eqnarray}
where
\begin{eqnarray}
Y_t(\lambda)
& := &
\tr\{ \rho\, V_\lambda A(t)\}
\, ,
\label{31}
\\
R_t
& := &
\!\int_0^\beta\! dx \!\int_0^x \! dy \, \tr\left\{\rho V_x V_y \psi(y) A(t)\right\}
\ .
\label{32}
\end{eqnarray}
Assuming temporarily
that 
\begin{eqnarray}
V_{th}:=\tr\{\rho V\}=0
\label{33}
\end{eqnarray}
(see also (\ref{4})), and omitting terms of order $g^2$,
we thus can rewrite (\ref{23}) by means of 
(\ref{28}) and (\ref{30}) as
\begin{eqnarray}
\langle A\rangle_t 
& = & 
\tr\{\rho A(t)\} + g \int_0^\beta d\lambda\,  Y_t(\lambda)
\ .
\label{34}
\end{eqnarray}

One readily infers from (\ref{1}), (\ref{4}), and (\ref{6}) that 
\begin{eqnarray}
\tr\{\rho A(t)\}=\tr\{\rho A\}=A_{th}
\ ,
\label{35}
\end{eqnarray}
yielding
\begin{eqnarray}
\Delta(t):=
\langle A\rangle_t 
- A_{th}=  g \int_0^\beta d\lambda\,  Y_t(\lambda)
\label{36}
\end{eqnarray}
up to corrections of order $g^2$.
Since the left hand side is real, one expects that also the
right hand side is real separately in every order of $g$.
In accordance with this expectation, one can explicitly
verify that the leading order term on the right hand side 
of (\ref{36}) is indeed real by exploiting that (\ref{31}) 
implies $Y_t^\ast(\lambda)=Y_t(\beta-\lambda)$.

Denoting the eigenvalues and eigenvectors of 
$H$ as $E_\nu$ and $|\nu\rangle$, respectively,
the corresponding matrix elements for instance of $V$ 
are abbreviated as 
\begin{eqnarray}
V_{\nu\mu}:=\langle \nu|V|\mu \rangle
\ .
\label{37}
\end{eqnarray}
Hence $\rho$ from (\ref{1}) becomes a diagonal matrix ,
\begin{eqnarray}
\rho_{\nu\mu} = \delta_{\nu\mu} p_\nu
\label{38}
\end{eqnarray}
with diagonal elements 
\begin{eqnarray}
p_\nu:=Z^{-1}e^{-\beta E_\nu}
\ .
\label{39}
\end{eqnarray}
Employing (\ref{1}), (\ref{6}), and  (\ref{25}) we thus 
can rewrite (\ref{31}) as
\begin{eqnarray}
Y_t(\lambda)
& = &
\tr\{\rho\, e^{\lambda H} V e^{-\lambda H} e^{iHt/\hbar} A e^{-iHt/\hbar}\}
\nonumber
\\
& = &
\sum_{\nu\mu}  p_\nu\,  e^{\lambda E_\nu} V_{\nu\mu} e^{-\lambda E_\mu} e^{i E_\mu t/\hbar} A_{\mu\nu} e^{-i E_\nu t/\hbar}
\nonumber
\\
& = &
\sum_{\nu\mu}  p_\nu\,  V_{\nu\mu} A_{\mu\nu} e^{E_{\mu\nu}(it/\hbar-\lambda)}
\ ,
\label{40}
\end{eqnarray}
where
\begin{eqnarray}
E_{\mu\nu}:=E_\mu-E_\nu
\ .
\label{41}
\end{eqnarray}
The last factor in (\ref{40}) can be rewritten as
\begin{eqnarray}
e^{E_{\mu\nu}(it/\hbar-\lambda)}
& = &
e^{i E_{\mu\nu} t/\hbar}\sum_{k=0}^\infty \frac{(-\lambda E_{\mu\nu})^k}{k!}
\nonumber
\\
& = &
\sum_{k=0}^\infty \frac{(-\lambda)^k}{k!}
(-i\hbar)^k\frac{d^k}{dt^k}e^{i E_{\mu\nu} t/\hbar}
\ ,
\label{42}
\end{eqnarray}
implying 
\begin{eqnarray}
Y_t(\lambda)
 = 
\sum_{k=0}^\infty \frac{(i\lambda\hbar)^k}{k!}\frac{d^k}{dt^k}
\sum_{\nu\mu}  p_\nu\,  V_{\nu\mu} A_{\mu\nu} e^{i E_{\mu\nu} t/\hbar}
\ .
\label{43}
\end{eqnarray}
Similarly as in (\ref{40}), the last double sum can be identified with $\tr\{\rho\, V\! A(t)\}$,
while the remaining integral over $\lambda$ in (\ref{36}) can 
now be carried out, yielding
\begin{eqnarray}
\Delta(t)
& = &  
g \beta \sum_{k=0}^\infty  \frac{(i\tau_B)^k}{(k+1)!}\,  
\frac{d^k}{dt^{k}}
\tr\{\rho\, V A(t)\}
\ .
\label{44}
\end{eqnarray}

So far, our derivation only applies for $V$'s with the property
$V_{th}=0$, see (\ref{33}).
The generalization to arbitrary $V$'s is straightforward:
To begin with, we define $\tilde V:=V- V_{th}$
and observe that employing $\tilde V$ instead of $V$ in (\ref{11})
does not affect any physically relevant system properties
such as the relations (\ref{1})-(\ref{10}) or (\ref{12})-(\ref{14}).
In particular, $\Delta (t)$ as defined in (\ref{36}), 
remains unchanged.
On the other hand,  $\tilde V$ now exhibits the property
$V_{th}=0$ and hence (\ref{44}) must apply
with $\tilde V$ instead of $V$.
Altogether, the generalization to arbitrary $V$'s
thus simply amounts to replacing $V$ in (\ref{44})
by $V- V_{th}$.
Exploiting (\ref{5}) and (\ref{35}), this finally yields (\ref{16}).

Moreover,  according to the 
discussion below Eq.~(\ref{36}) 
the right hand side of
(\ref{44}) and thus of (\ref{16}) 
must be a purely real function of $t$,
as claimed below Eq. (\ref{17}).

\section{Validity of the approximation}
\label{s3}

Combining (\ref{23}), (\ref{28}), (\ref{30}), (\ref{33}),
and (\ref{35})
yields
\begin{eqnarray}
\langle A\rangle_t 
= \frac{A_{th}+  g \int_0^\beta d\lambda \tr\{\rho V_\lambda A(t)\} + g^2 R_t}{1
+g^2 Q}
\ .
\label{45}
\end{eqnarray}
Our main (and only) approximation in 
Sec. \ref{s2}
was to neglect the terms of order $g^2$ in (\ref{45}),
resulting in (\ref{34}).
However, since 
$Q$ and $R_t$ 
in (\ref{45})
actually still depend on $g$ themselves 
via 
(\ref{29}), (\ref{32}),
and (\ref{20}),
one may feel that a more careful justification of 
such an approximation
would be desirable.
This is 
accomplished
in Appendix~\ref{app1}
in the form of rigorous upper bounds 
for $|Q|$ and $|R_t|$.
Most importantly, and in contrast to ordinary
time-dependent perturbation theory, those bounds
are not restricted to sufficiently small times
but rather apply uniformly in $t$.

{The derivation of those bounds 
is based on the following two premises: 
The system Hamiltonian $H$ must be
a sum of local operators, and the perturbation
$V$ must be a local operator,
see Appendix \ref{app1} for further details.
From now on, these two premises are 
thus tacitly taken for granted.}

{Roughly speaking, the relevant small parameter
in the detailed analytical considerations in 
Appendix \ref{app1} is given, as one might 
have intuitively expected, by 
$g\beta\opnorm{V}$
(operator norm):
If this parameter is small, the 
neglected corrections (of higher order in $g$)
on the right hand side of (\ref{16})
and (\ref{18}) can be shown to be 
small.
More precisely speaking, also the system Hamiltonian
$H$ must of course somehow enter the game,
giving rise to an extra factor $f(\beta)$
(see Eq. (\ref{a33})).
The latter accounts for the specific properties of 
$H$ as detailed below 
Eq. (\ref{a33}), and is often expected to be of the 
order of unity for small-to-moderate $\beta$ values
when working in natural units.}

Another very important question is how  
$Q$ and $R_t$ 
in (\ref{45}) depend on the size of 
the considered system, and, in particular, how 
they behave in the thermodynamic limit \cite{f0}.
This issue is clearly far from obvious in view of 
(\ref{29}), (\ref{32}), and the exponential dependence 
on the system Hamiltonians in (\ref{6}), (\ref{20}), 
and (\ref{25}).
From the rigorous considerations in Appendix \ref{app1} one can 
infer the following quite strong conclusion regarding this issue:
If $\opnorm{V}$ and $\opnorm{A}$ 
remain bounded in the thermodynamic limit, 
then the same must apply to 
$|Q|$ and $\max_t|R_t|$.
Therefore,
$Q$ and $R_t$ are indeed negligible in (\ref{45})
for sufficiently small $g$ even after taking 
the thermodynamic limit.

{Due to our above premise that $V$ is a local
operator, the requirement that  $\opnorm{V}$ must 
remain bounded in the thermodynamic limit is
(almost) automatically fulfilled.
Similarly, since the entire formalism is 
linear in $A$, and since most physically relevant
observables are (sums of) local operators,
the requirement that  $\opnorm{A}$ must remain 
bounded in the thermodynamic limit does not 
amount to any substantial loss of generality either.}

{We finally note that the extension of our present analytical
results to non-local perturbations $V$ amounts to a very challenging
unsolved problem, as can be understood by} 
the following non-rigorous arguments.
Let us, for instance, consider a local observable $A$ and an
extensive perturbation $V$
(meaning that $V$ consists of a sum of local operators
which grows extensively with the system size \cite{f0}).
For physical reasons it then seems intuitively reasonable to expect
that the quantities  $\langle A\rangle_t$ and $A_{th}$
in (\ref{45}) exhibit a well-defined (finite) thermodynamic 
limit \cite{f1}.
Turning to the second term in the numerator of (\ref{45}),
we observe that this term 
coincides with the right hand side of (\ref{36}) and thus with (\ref{44}) and (\ref{16}).
As before, 
it is
reasonable to expect
that the correlations $C_{\! V\!\!A}(t)$ in
(\ref{5}) exhibit a well-defined (finite) 
thermodynamic 
limit \cite{f2}.
Hence, also (\ref{16}) is expected to remain finite in 
the thermodynamic limit.
Finally, by means of numerical explorations 
and of similar analytical considerations 
as in Appendix \ref{app1} we found that
both $Q$ and $R_t$ 
generally seem to diverge in the
thermodynamic limit.
[Incidentally, $R_t/Q$ must still converge to $\langle A\rangle_t$
according to (\ref{45}).]
Thus, neglecting $Q$ and $R_t$ in (\ref{45})
can only be justified for smaller and smaller $g$-values 
as the system size increases.
In other words, the main approximation 
of our present approach
breaks down for any finite $g$ in the thermodynamic limit.

\section{General remarks}
\label{s4}

Although systems with many degrees of freedom (DOF's) are usually of foremost
interest (see also Sec.~\ref{s3}), it is nevertheless 
remarkable that 
our main results actually apply to systems with an arbitrary number 
of DOF's, including very ``small'' systems.

Similarly, it is 
remarkable that  issues like integrability, ergodicity, 
or many-body localization do not 
play any role.

In most previous theoretical works in this context,
the physical interest of
such temporal correlations as in (\ref{5}), (\ref{7}), (\ref{8})
is considered as self-evident.
Here, we adopt the same viewpoint that relations 
like (\ref{15})-(\ref{18}) are of considerable 
theoretical interest in themselves 
without any further discussion of whether and how
such temporal correlations may be experimentally 
measurable.
[The latter is quite obvious in the special case $t=0$,
which we address in more detail below, but not any more
for $t\not=0$.]
For a more detailed discussion of 
such issues
we refer to \cite{tal86,shi17,alh20} and further references therein.

As already mentioned below (\ref{19}), Onsager's original hypothesis 
(\ref{15}) is recovered for asymptotically small values of  
$\tau_B$ in (\ref{18}).
Moreover, one can infer from (\ref{18}) that the deviations from
Onsager's hypothesis remain small as 
long as the characteristic time scale of the correlations 
in (\ref{5}) is large compared to the Boltzmann 
time $\tau_B$
\cite{f3}.
Experimentally, this will be often the case unless we are 
dealing with 
rather
low temperatures.
On the other hand, in numerical studies, where choosing units with
$\hbar=1$ is quite natural, 
those corrections are expected to be 
non-negligible in many cases 
(see also Sec. \ref{s5} below).

For sufficiently large systems
one often expects, observes, or even can show \cite{equil}
that the time-dependent expectation values in (\ref{13})
approach a (nearly) constant value 
after initial transients have died out,
and analogously for the the temporal 
correlations in (\ref{5}) \cite{alh20}.
{Quantitatively,} 
these constant values can be 
{determined}
by taking the
long-time average on both sides of (\ref{18}).
One readily sees that this 
average is zero for each summand with $k>0$ 
on the right hand side of (\ref{18}).
Remarkably, the dependence on $\tau_B$ thus
disappears altogether
and we recover once again the same result as in
Onsager's original relation (\ref{15})
as far as long-time averages are concerned.

On the other hand, for large but finite systems one expects that
both sides in (\ref{18}) still exhibit (even for arbitrarily late times) 
some small temporal fluctuations around their 
long-time averages \cite{alh20,equil}.
A quite non-trivial prediction of our present work
is that those long-time fluctuations
are again connected via the relation (\ref{18}).
Furthermore, it is not obvious at all
how significant the finite $\tau_B$ corrections of Onsager's 
hypothesis (\ref{15}) will be with respect 
to those long-time fluctuations.

More generally speaking, our present theory must remain
valid even in such ``unusual'' cases where the system
does not exhibit any kind of relaxation, 
equilibration, or thermalization
in the long run. For instance, this may be the
case for systems with few DOF's or due to some
special symmetries and conservation laws.

Yet another interesting special case arises for $t=0$.
Namely, the left hand side in the relations
(\ref{15}), (\ref{16}), and (\ref{18}) 
then amounts to the difference between the thermal 
expectation values for two sightly different
systems with Hamiltonians $H_g$ and $H$
(see Eqs. (\ref{1}), (\ref{4}), and  (\ref{9})-(\ref{13})).
In other words, all quantities appearing in
(\ref{15}), (\ref{16}), and (\ref{18}) 
now solely refer to thermal equilibrium properties.
In the context of equilibrium statistical mechanics, 
such 
equations
are sometimes also denoted for instance as 
fluctuation-response relations,
though the choice of this and other names 
tends to be
somewhat
unfortunate 
\cite{shi17}.
While Eq. (\ref{15}) with $t=0$ amounts to a textbook
relation of this kind, our present results
(\ref{16}) and (\ref{18}) amount to a non-trivial
improvement for quantum systems, which
to our knowledge has not been previously known.

\section{Quantitative Examples}
\label{s5}

Here, we will 
quantitatively
illustrate some of the general 
issues addressed in the previous 
sections by means of specific examples.
A systematic exploration of all those various 
analytical predictions is clearly impossible,
hence we will confine ourselves to just a few 
instances.

In particular, we will {mainly} focus on one-dimensional 
Heisenberg-type Hamiltonians 
(XYZ models)
with open boundary conditions,
\begin{eqnarray}
H & = & -
\!\!\!\!
\sum_{a\in\{x,y,z\}} 
\!\!\!\!
J_a  \sum_{l=1}^{L-1} s^a_{l+1}s^a_l + \, h  \sum_{l=1}^{L} s^x_l
\, ,
\label{46}
\end{eqnarray}
where $s^{x,y,z}_l$ are spin-1/2 operators at
the lattice sites $l\in\{1,...,L\}$, the 
$J_{x,y,z}$ quantify the coupling strength 
of the nearest-neighbor interactions,
and $h$ may be considered to account for an 
external magnetic field.
Focusing on non-vanishing 
and non-identical
couplings $J_{x,y,z}$,
this model is non-integrable for $h\not=0$
\cite{dmi02},
and integrable (but still of a non-trivial, so-called interacting type)
for $h=0$ \cite{skl88,alc87,bei13}.
Moreover, such models are commonly expected to 
exhibit a well-defined thermodynamic limit
when $L\to\infty$ (see Sec. \ref{s3}).

As mentioned 
{in} Sec. \ref{s3},
the perturbation $V$ in (\ref{11}) is assumed to 
be a local operator.
As a particularly simple choice 
we will mainly consider examples of the form
$V=s_l^z$ for some $l\in\{1,...,L\}$.
Physically, it is reasonable to expect 
(and numerically seen) 
that such a local perturbation will only
lead to a notable response for observables 
which are not too far away from the perturbation,
and will be particularly well visible  
if the observable is in some sense 
``similar''  to the perturbation.
Therefore, it is natural to focus on 
observables of the form $A=V$.

Yet another appealing feature of
the above choice $A=V=s_l^z$
is that the corresponding thermal expectation 
values $A_{th}$ in (\ref{5}) can be shown 
analytically to vanish for symmetry reasons 
\cite{eid23}.
Moreover, one readily concludes that
$VA(0)=(s_l^z)^2=1/4$.
Together with (\ref{5}) it follows that
$C_{\! V\!\!A}(0)=1/4$.
Exploiting (\ref{7}), Onsager's regression 
hypothesis at the time-point $t=0$ thus
assumes the 
simple form
\begin{eqnarray}
\langle A\rangle_{t=0} = g\beta/4
\label{46a}
\end{eqnarray}
independent of any further details of the considered system.
Since the left hand side in (\ref{46a}) is bounded by the largest and 
smallest eigenvalues of $A=s_l^z$, that is, by $\pm1/2$, 
the prediction (\ref{46a}) 
will certainly be wrong
when $|g\beta|>2$.
To the best of our knowledge, this is the simplest
analytical example for the failure of Onsager's regression 
hypothesis in the quantum regime.

Our next observation is
that a spin-1/2 model like in (\ref{46}) does not 
admit a physically meaningful classical limit.
The basic reason is that the classical limit requires
the emergence of an asymptotically continuous 
level density for $\hbar\to 0$
(the simplest example is a harmonic oscillator),
while the number of energy levels in 
(\ref{46}) is finite and independent of $\hbar$.
As usual in numerical explorations of such models, 
we thus work in units with
\begin{eqnarray}
\hbar =1
\ ,
\label{47}
\end{eqnarray}
implying that $\tau_B$ in (\ref{17}) formally coincides with $\beta$.

The general strategy in our subsequent numerical explorations
will be to 
compute
the left hand side in (\ref{18}) and compare it with the numerically evaluated 
functions
\begin{eqnarray}
P_K(t) := g \,\sum_{k=0}^K \frac{\beta^{k+1}}{(k+1)!} \,S_k(t)
\, ,
\label{48}
\end{eqnarray}
converging towards the right hand side of (\ref{18}) 
for large values of $K$, while $K=0$ corresponds to 
Onsager's hypothesis (\ref{15}).

In passing we note that the dependence of (\ref{48}) on $\beta$ 
is not as simple as it might appear at first glance
since also the functions $S_k(t)$ depend on $\beta$ 
according to (\ref{1})-(\ref{8}) and (\ref{19}).
Nevertheless, it is reasonable to expect, and will be later 
confirmed numerically, that 
the factor $\beta^{K+1}$ is asymptotically dominating in (\ref{48}) 
for large $\beta$.

In the following subsections we will present
our results for the left hand side 
of (\ref{18}) and for the functions $P_K(t)$ in (\ref{48}).
These results have been obtained by the numerically exact
diagonalization of the pertinent Hamiltonian
$H$ in (\ref{46}).
Moreover, we have circumvented the numerical evaluation 
of the $k$-th derivative
appearing in (\ref{19}) 
as follows:
As a first step, we observe that  (\ref{1}) and (\ref{6}) imply
\begin{eqnarray}
\tr\{ \rho \, V(s) A(t)\}=\tr\{ \rho \, V A(t-s)\}
\label{49}
\end{eqnarray}
for arbitrary $t,s\in\RR$, 
and hence
\begin{eqnarray}
& & \tr\{\rho\, V^{(k)}(0) A(t)\}
=  \frac{d^k}{ds^{k}}
\tr\{\rho\, V(s) A(t)\}\big|_{s=0} =
\nonumber
\\
& & 
=
\frac{d^k}{ds^{k}}
\tr\{\rho\, V A(t\!-\!s)\}\big|_{s=0}
=
(-1)^k\,
\tr\{\rho\, V A^{(k)}\!(t)\}
\, . \ \ \ \ \ \
\label{50}
\end{eqnarray}
As a second step,
we define (as usual) $\dot V:= i\, [H,V]$ (see also (\ref{47})),
while the higher derivatives $V^{(k)}$ then follow
recursively as
\begin{eqnarray}
V^{(k+1)}:= i\, [H,V^{(k)}]
\ .
\label{51}
\end{eqnarray}
Alternatively, if we set $V^{(0)}:=V$ then (\ref{51})
is valid for all $k\in\NN_0$.
As expected, it follows by exploiting (\ref{6}) that
\begin{eqnarray}
V^{(k)}(0) = V^{(k)}
\ ,
\label{52}
\end{eqnarray}
and thus with (\ref{5}) and (\ref{50}) that
\begin{eqnarray}
C^{(k)}_{\! V\!\!A}(t) = (-1)^k \,\tr\{\rho\, V^{(k)} A(t)\} 
\label{53}
\end{eqnarray}
for all $k\geq 1$. Analogous relations apply to the derivatives
of the real and imaginary parts in (\ref{7}), (\ref{8}), which 
arise in (\ref{19}) and  are thus needed in (\ref{48}).
The main point is that the commutators, appearing
on the right hand side of (\ref{53}) via (\ref{51}), are numerically 
much more convenient and accurate than directly
evaluating the $k$-th derivative 
on the left hand side of (\ref{53}).

\subsection{Numerical results for $t=0$}
\label{s51}

As mentioned at the end of Sec. \ref{s4}, in the special case $t=0$ 
our findings are essentially tantamount to a generalization of the 
so-called fluctuation-response relation
in the context of equilibrium 
statistical mechanics.

\begin{figure}
\hspace*{-0.8cm}
\includegraphics[scale=0.95]{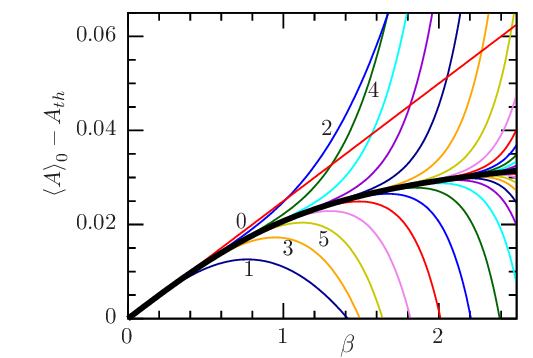}
\caption{Numerical results for the non-integrable 
spin-model from (\ref{48})
with $L=16$, $J_x=1$, $J_y=1.2$, $J_z=1.5$, $h=1$,
and a local perturbation in (\ref{11}) with $g=0.1$ and $V=s^z_{L/2}$.
Bold black line:  Left hand side of (\ref{18}) versus $\beta$ for $A=s^z_{L/2}$ and $t=0$.
Colored lines: Numerical results for $P_K(0)$ from (\ref{48}) 
with $K=0,1,...,30$.
The first few $K$-values are indicated as numbers close to the 
corresponding lines. 
The continuation for all other $K$ is obvious.
In particular, the straight red line ($K=0$) represents Onsager's 
regression hypothesis from (\ref{15}).
}
\label{fig1}
\end{figure}

\begin{figure}
\hspace*{-0.8cm}
\includegraphics[scale=0.95]{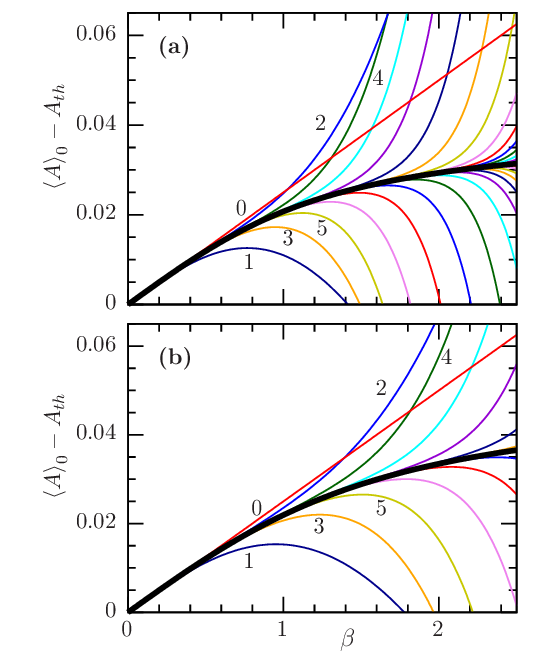}
\caption{Same as in Fig. 1 but for smaller systems with 
$L=8$ in (a) and $L=2$ in (b).
}
\label{fig2}
\end{figure}

\begin{figure}
\hspace*{-0.8cm}
\includegraphics[scale=0.95]{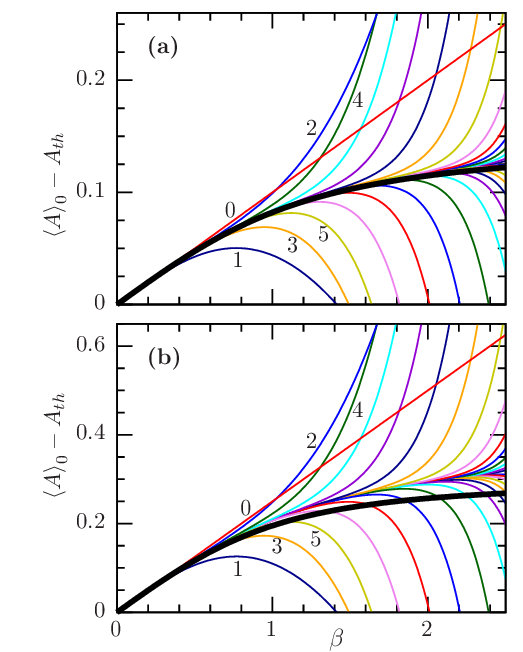}
\caption{Same as in Fig. 1 but for stronger perturbations with
$g=0.4$ in (a) and $g=1$ in (b).
}
\label{fig3}
\end{figure}

\begin{figure}
\hspace*{-0.8cm}
\includegraphics[scale=0.95]{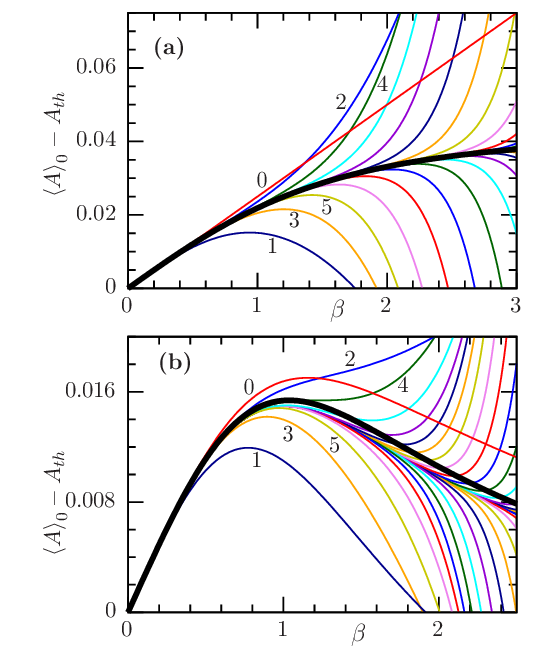}
\caption{Same as in Fig. 1 but for a different perturbation 
and observable,
{namely $V=A=s_{1}^z$ in (a) and  $V=A=s_{L/2}^x$ in (b).}
Moreover,  {in (a)} the depicted range of $\beta$-values 
is somewhat larger than in Fig. \ref{fig1}.
}
\label{fig4}
\end{figure}

\begin{figure}
\hspace*{-0.8cm}
\includegraphics[scale=0.95]{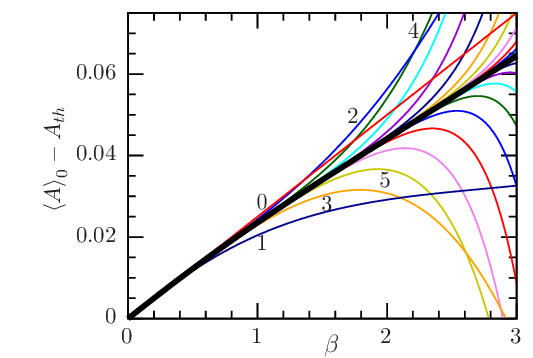}
\caption{Same as in Fig. 1 but for an integrable model
with $h=0$ in (\ref{48}).
Moreover, the depicted range of $\beta$-values 
is somewhat larger than in Fig. \ref{fig1}.
}
\label{fig5}
\end{figure}

A first example is depicted in Fig. \ref{fig1}.
The main conclusion is that Onsager's regression 
hypothesis (straight red line, $K=0$) indeed becomes 
quite bad for large $\beta$ (low temperatures),
while our improved prediction from (\ref{18}) 
(corresponding to $K\to\infty$) works very well.
Furthermore, and as announced below Eq. (\ref{48}), 
for any given (finite) $K$ value, the function $P_K(0)$
indeed seems to diverge 
{approximately as $\beta^{K+1}$ for large $\beta$.
Some interesting additional features, which are numerically observed,
and which can also be to some extent understood as a consequence 
of this divergence and the fact that the left hand side of (\ref{18}) 
is bounded (uniformly in $\beta$),
are as follows:
(i)
Upon increasing $K$, the functions $P_K(0)$ from (\ref{48})
diverge for asymptotically large $\beta$ alternatingly
towards plus and minus infinity.
(ii) 
The convergence of the finite sums
$P_K(0)$ from (\ref{48}) towards the infinite sum 
in (\ref{18})} cannot be uniform in $\beta$:
The larger the value of $\beta$, the larger values of 
$K$ are needed until convergence sets in.
(iii) For sufficiently large but fixed $\beta$, the approximations
$P_K(0)$ considered as a function of $K$
will initially (for small-to-moderate $K$) even become 
worse upon increasing $K$, and only later 
(for sufficiently large $K$) converge towards a 
really good approximation of the 
{infinite sum in (\ref{18}).}

We also remark that the perfectly straight numerical lines for 
$K=0$ in Figs. \ref{fig1}-\ref{fig5}
are in agreement with the
analytical prediction from (\ref{46a}).
{An exception is Fig. \ref{fig4}(b),
where the requirement $A=V=\sigma_l^z$ above Eq. (\ref{46a})
is not fulfilled,  and hence no straight line arises
for $K=0$.}

Turning to Fig. \ref{fig2},
its main message (together with Fig. \ref{fig1}) is
that the dependence 
on the system size $L$ is remarkably weak.
In other words, the thermodynamic limit is approached 
very quickly upon increasing $L$.
We conjecture that our employing canonical 
ensembles as initial states in (\ref{9}) 
may play an important role in this context, 
since similar observations have also been 
reported in various other numerical 
explorations in the literature.
We also see that, as predicted in Sec. \ref{s3}, 
our {analytical} theory indeed works very well independently
of whether the considered system is small or large.

Fig. \ref{fig3} exemplifies the 
impact of
the neglected higher order terms in $g$.
In agreement with the analytics at the end of Appendix \ref{app1}
{(see also Sec. \ref{s3}),
the essential}
quality parameter for the reliability of  
our present
linear response type approximation is the product 
{$g\beta\opnorm{V}$,
where $\opnorm{V}=1/2$ 
in all our numerical examples.}
Quantitatively, the differences between the numerically
exact results (bold black lines) and the analytical linear 
response theory (colored lines with sufficiently large $K$)
indeed become 
visible on the scale of these plots when $\beta$ 
is comparable or larger than $1/g$.

Fig. \ref{fig4} illustrates the behavior for a different choice of the
perturbation $V$ and the observable $A$.
{Particularly noteworthy are the following features of
Fig.~\ref{fig4}~(b):
 (i)~Quantitatively, the observed effects are generally 
 weaker than in the other examples (note the $y$-axis scales).
 (ii) The relative deviations (finite-$g$ effects)
 between the numerically exact results (black)
 and our analytics (corresponding to $K\to\infty$) are larger
 than in 
 Figs. \ref{fig1}, \ref{fig2}, \ref{fig4} (a),
 but the absolute deviations remain 
 comparable.
 (iii) Unlike in Fig. \ref{fig3}, those deviations hardly grow with
 increasing $\beta$ as far as their absolute values are concerned,
 but the relative deviations still grow in nearly the same way.
 (iv) Similar considerations apply to the differences between 
 Onsagers's hypothesis ($K=0$, red) and the numerically exact
 results.
 (v) As announced, Onsager's hypothesis no longer amounts to 
 a straight line.}
 
For the rest, it does not seem to us of great interest 
to present here further results for still other 
examples of $V$ and $A$, see also the general 
considerations above Eq. (\ref{46a}).

{Turning to Fig. \ref{fig5}, a comparison of these 
results (for $h=0$) 
with those in the other examples (with $h=1$)}
confirms the prediction from Sec. \ref{s4}
that our {analytical} theory works very well for non-integrable
($h\not=0$) as well as for integrable ($h=0$) models.

\begin{figure}
\hspace*{-0.8cm}
\includegraphics[scale=0.95]{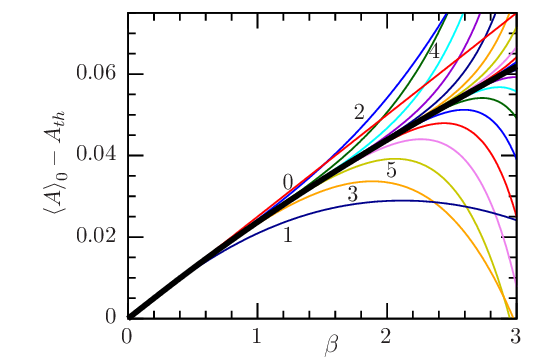}
\caption{{Same as in Fig. 1 but for the frustrated spin chain model
from (\ref{46b}) with $J=0.25$.
Moreover, the depicted range of $\beta$-values 
is somewhat larger than in Fig. \ref{fig1}.}
}
\label{fig6}
\end{figure}

{As yet another, somewhat more ``exotic'' example, we finally 
consider a so-called frustrated ferromagnetic Heisenberg 
spin-$1/2$ chain of the form \cite{rot98,sud09}
\begin{eqnarray}
H & = &  - \sum_{l=1}^{L-1} \vec s_{l+1}\cdot \vec s_l 
+ J \sum_{l=1}^{L-2} \vec s_{l+2}\cdot \vec s_l 
\label{46b}
\end{eqnarray}
with $J>0$ (frustration) and $\vec s_l:=(s_l^x,s_l^y,s_l^z)$, 
see also below Eq. (\ref{46}).
Besides the special feature of frustration, this model is also 
known to exhibit numerous degeneracies (due to 
its $SU(2)$ symmetry) and  to be non-integrable.
The concomitant numerical findings in Fig. \ref{fig6}
are quite similar to those in (\ref{fig5}),
indicating that frustration and/or degeneracies
do not seem 
to play an important role
with respect to the
questions in which we are interested here.}

\subsection{Time-dependent numerical results}
\label{s52}

\begin{figure}
\hspace*{-0.8cm}
\includegraphics[scale=0.95]{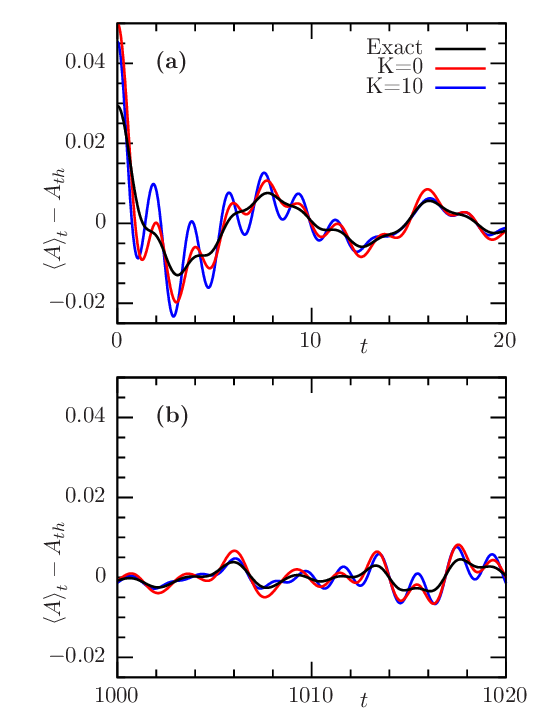}
\caption{
Numerical results 
for the same model as in Fig.~1
with $\beta=2$.
Black lines:  Left hand side of (\ref{18}) versus $t$
for (a) $t\in[0,20]$ (initial behavior) and (b) $t\in[1000,1020]$ 
(long-time behavior).
Colored lines: Numerical results for $P_K(t)$ from (\ref{48}) with
$K=0$ (red) and $K=10$ (blue).
The corresponding results 
$K=20$ and $K= 30$ were found to be
indistinguishable from the 
black lines and are therefore not shown.
}
\label{fig7}
\end{figure}

\begin{figure}
\hspace*{-0.8cm}
\includegraphics[scale=0.95]{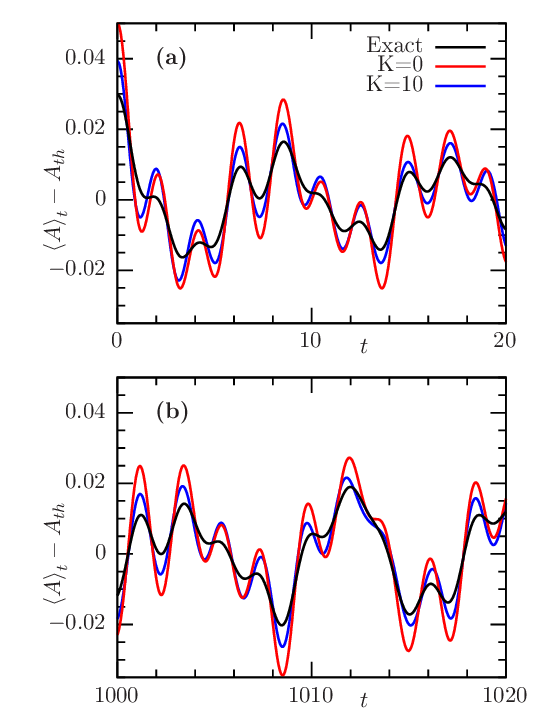}
\caption{Same as in Fig. \ref{fig7} but for  a smaller system with $L=4$.
}
\label{fig8}
\end{figure}
 
Fig. \ref{fig7} depicts the time-dependent comparison of the
left hand side in Eq. (\ref{18}) with the approximations on the
right hand side in Eq. (\ref{48}) for two representative $K$-values.
As before, Onsager's regression hypothesis (red lines, $K=0$) 
quite notably deviates from the numerically exact results (black).
The approximation for $K=10$ (blue) is sometimes better and 
sometimes worse than 
that for $K=0$.
This 
is in accordance with the fact that the lines
for $K=0$ and $K=10$ in Fig. \ref{fig1} cross each other 
near $\beta=2$ 
(see also item (iii) in Sec.~\ref{s51}).
{For $K\geq 20$, our numerical data} 
were found to practically
coincide with the 
black lines
in Fig. \ref{fig7}. 
In other words, our 
{analytical
theory (corresponding to $K\to\infty$) indeed agrees}
very well with the numerically
exact results 
for all times $t$ which we actually explored.

We also observe that the finite-$K$ effects in Fig. \ref{fig7}
are particularly pronounced when the numerically exact 
curves (black) exhibit large curvatures,
and are comparatively small near turning 
points.
Unfortunately, we did not succeed in coming up with a simple intuitive 
explanation of this observation.

While Fig. \ref{7}(a) illustrates the initial relaxation 
behavior for $t\in[0,20]$, the everlasting long-time 
fluctuations discussed at the end of Sec. \ref{s4} 
are exemplified by the behavior for $t\in[1000,1020]$ in
Fig. \ref{fig7}(b).
Though the detailed behavior for those large times is quite non-trivial,
the agreement with our {analytical} theory is
still nearly perfect for sufficiently large $K$, while the
deviations for $K=0$ (Onsager's hypothesis) are still quite
notable.

Finally, Fig. \ref{fig8} confirms 
(upon comparison with Fig.~\ref{fig7})
the theoretical prediction \cite{equil,alh20} (see also Sec. \ref{s4})
that these long-time fluctuations 
quickly decrease as the system size is increased.
For the rest (initial relaxation behavior, long-time average), 
the dependence on the system size is 
rather
weak, 
as already observed in the previous subsection.
Moreover, Fig. \ref{fig8} once more confirms the 
prediction that {our analytical} theory works 
very well {even} for small systems and/or
in the absence of  {a distinct ``initial relaxation''.} 

\section{Generalization}
\label{s6}

Our goal is to extend the results from Sec. \ref{s1} to
more general density operators than in (\ref{1}) and (\ref{9})
provided the system satisfies certain additional assumptions.

Namely, we assume that there exists a projector $P$ 
onto some sub-Hilbert space $\tilde\hr$ with the property
\begin{eqnarray}
[H,P]=[V,P]=0
\ .
\label{55}
\end{eqnarray}
The most important example arises when the system Hamiltonian
$H$ as well as the perturbed Hamiltonian $H_g$ from (\ref{11})
exhibit a common conserved quantity
$S$, i.e.,
\begin{eqnarray}
[H,S]=[V,S]=0
\ ,
\label{56}
\end{eqnarray} 
and if we identify $\tilde\hr$ with one of the eigenspaces 
of $S$.
(One readily verifies that the property (\ref{55}) is indeed
fulfilled in such a case.)
More generally, if there are several (commuting) conserved quantities
of $H$ and $H_g$,  the subspace $\tilde \hr$ may be chosen
as an eigenspace of one of those conserved quantities,
but also as a common eigenspace of several conserved quantities.

Given that (\ref{55}) is fulfilled, we show in Appendix \ref{app2} that our main 
results from (\ref{16})-(\ref{19}) are still valid if the canonical ensembles in
(\ref{1}) and (\ref{9}) are replaced by 
\begin{eqnarray}
\rho & := & P e^{-\beta H}/\, \tr\{P e^{-\beta H}\}
\ ,
\label{57}
\\
\rho_0 & := & P e^{-\beta H_g}/\, \tr\{P e^{-\beta H_g}\}
\ .
\label{58}
\end{eqnarray}
In other words, we can
replace the original 
density operators from (\ref{1}) and (\ref{9}) by their
projections/reductions onto any invariant subspace $\tilde \hr$,
whereas the observable $A$ and the system Hamiltonian $H$
still remain entirely unchanged for instance in (\ref{4})-(\ref{6}) and in 
(\ref{12})-(\ref{14}).

While one may come up with various intuitive arguments of why these
findings might not be entirely unexpected, the details of a more rigorous
line of reasoning are not obvious at all, see Appendix \ref{app2}.

If relations as in (\ref{55}) apply simultaneously to two different 
projectors $P_1$ and $P_2$,
our main results from (\ref{16})-(\ref{19}) are valid for each of the
two corresponding ensembles of the form (\ref{57}) and (\ref{58}).
It might thus seem tempting to conjecture that also linear combinations
thereof are still be admissible.
However, this cannot be true in view of the fact that
the left hand side of (\ref{16}) is linear in $\rho$, while
the last term in (\ref{5}) and thus the right hand side of 
(\ref{16}) is non-linear in $\rho$.

\begin{figure}
\hspace*{-0.8cm}
\includegraphics[scale=0.95]{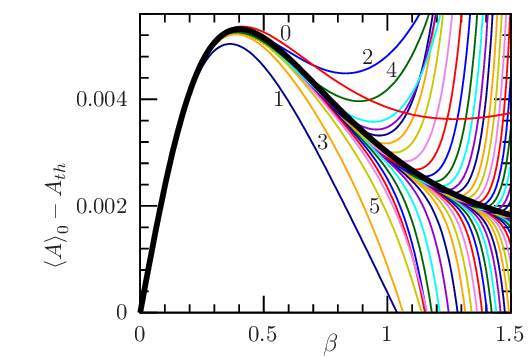}
\caption{{Numerical results for the random field spin chain from (\ref{46c})
with disorder strength $h_{max}=5$
for an observable $A$ and a perturbation
in (\ref{11}) with $A=V=s^x_{L/2}$ and $g = 0.1$.
Furthermore, Eqs. (\ref{57}) and (\ref{58}) have been employed
with a projector $P$ onto the eigenspace of the conserved
quantity from (\ref{46d}) with vanishing eigenvalue 
(subsector with zero total magnetization).
All further details are as in Fig. \ref{fig1} except that 
curves up to $K=50$ are shown.}
}
\label{fig9}
\end{figure}

In order to numerically illustrate these analytical findings, we
consider a random field Heisenberg spin-$1/2$ chain of the form
\begin{eqnarray}
H & = &  \sum_{l=1}^{L-1} \vec s_{l+1}\cdot \vec s_l 
+\sum_{l=1}^L h_l s_l^x
\ ,
\label{46c}
\end{eqnarray}
where $\vec s_l:=(s_l^x,s_l^y,s_l^z)$, see also below Eq. (\ref{46}).
Moreover, the $h_l$ are independent random variables,
uniformly distributed within the interval 
$[-h_{max},h_{max}]$ for some arbitrary 
but fixed 
$h_{max}\geq 0$.
This model is thus quite similar to (\ref{46}),
except that all nearest neighbor
coupling strengths $J_a$ are now chosen
equal to $-1$ and the external magnetic field 
is now randomized. 

The model (\ref{46c}) is one of the standard examples considered 
in the context of many-body localization (MBL),
see for instance Refs. \cite{pal10,nan15} and 
further references therein.
Regarding the disorder strength $h_{max}$, we 
focus on the choice $h_{max}=5$, for which the
model has been reported to exhibit MBL 
in Ref.~\cite{pal10}.
We remark that whether and in which sense 
this finding from \cite{pal10} is actually correct
is still debated in the literature.
This controversy is in itself not of immediate
relevance in our present paper:
For us, the model (\ref{46c}) just serves
the purpose to further enlarge the diversity of different
models we are numerically exploring.

Particularly important in our present context 
is the well-known fact \cite{pal10} that the total 
magnetization
\begin{eqnarray}
S:=\sum_{l=1}^Ls_l^x
\label{46d}
\end{eqnarray}
is a conserved quantity of the model from (\ref{46c}).
Similarly as in Ref. \cite{pal10}, we thus restrict ourselves
to the eigenspace $\tilde \hr$ of $S$ with eigenvalue zero,
and thus with the corresponding projector $P$ onto $\tilde\hr$ 
appearing in Eqs. (\ref{57}) and (\ref{58}).
Moreover, we focus on perturbations 
of the form $V=s^x_l$ in order to
fulfill the second condition in (\ref{55}).

The main qualitative features of our numerical findings
in Fig. \ref{fig9} are quite similar to those in Fig. \ref{fig4}(b),
suggesting that neither the disorder 
nor the restriction to a subspace of the 
conserved quantity seems to have a major
impact with respect to the main issues of our 
present paper.

\section{Conclusions}
\label{s7}

Our main result consists in a modification of Onsager's regression
hypothesis (see Eq. (\ref{18})), which 
{is}
analytically shown to be 
correct in linear order of the perturbation strength $g$ in (\ref{11}).
In contrast to most other perturbative approaches in similar 
contexts, the neglected non-linear corrections can be rigorously
bounded for a large class of physically relevant
model systems and for arbitrarily large times $t$
{(see Sec. \ref{s3})}.
Further noteworthy general features of the theory 
are collected in {Secs. \ref{s4} and \ref{s6}}, 
while more detailed quantitative examples are 
provided in {Secs. \ref{s5} and \ref{s6}.}

Since Onsager's regression hypothesis is known
to fail in the quantum regime \cite{tal86,for96,shi17}
(see also around Eq. (\ref{46a})),
a common proposal has been that the 
adequate
quantum version 
of the hypothesis
should be nothing else than the well-established fluctuation-dissipation 
theorem (FDT)
\cite{for96,bon12,els13}.
This proposal is incorrect for the following reasons.
With respect to the specific questions in which we are 
interested here, a very basic feature of the 
FDT is that it does not admit any non-trivial prediction 
concerning the initial time point $t=0$.
On the other hand, Onsager's regression hypothesis 
from (\ref{15}) does make a non-trivial prediction 
for $t=0$.
While the latter prediction is still not quantitatively 
right, it is appropriately corrected by our present 
theory (see Sec. \ref{s51}).

Formally speaking, Onsager's {original}
regression hypothesis 
works well in the combined linear response and 
near-classical regime, i.e.,
when both the perturbation parameter 
{$g\beta\opnorm{V}$ 
(see Sec. \ref{s3})}
and the Boltzmann time $\tau_B$ 
in (\ref{17}) are simultaneously small.
Or, when working in 
{natural}
units with $\hbar=1$
{and $\opnorm{V}=1$},
both $g\beta$ and $\beta$
must be simultaneously small.
In contrast, our present theory works well as long as
the product $g\beta$ is small, while 
{$\beta$
may not be small.
On the other hand,}
in order to explore very low temperatures 
{(large $\beta$)}
and not too small perturbations, one necessarily must 
go beyond the present linear response regime.
This will be the subject of a separate publication.

A quite common practice in cases like ours is to
devise some sort of expansion in the perturbation
parameter,
and then to simply neglect all 
non-linear terms
without any further consideration of their actual 
effect.
Such a strategy has been severely criticised 
in a hardly accessible but nevertheless highly cited paper 
by van Kampen \cite{kam71}, see also
Section 2 in Ref. \cite{bon12}.
Within our present approach, those 
non-linear terms can be rigorously bounded 
for a large class of 
physically relevant model systems.
On the other hand, we pointed out in Sec. \ref{s3} that the
higher order terms are seen to diverge
in the thermodynamic limit for other model classes,
most prominently when the perturbation operator
is an extensive quantity.
In other words, the widespread opinion that van Kampen's
criticism is irrelevant ``for all practical purposes''
may not be appropriate.
A more detailed account of our ongoing explorations
in the case of extensive perturbations will be given
elsewhere.

\vspace*{0.8cm}
\begin{acknowledgments}
We are indebted to Anatoly Dymarsky for 
informing us about 
his important rigorous results in Ref. \cite{avd20}
and about unpublished generalizations thereof.
This work was supported by the 
Deutsche Forschungsgemeinschaft (DFG, German Research Foundation)
under Grant No. 355031190 
within the Research Unit FOR 2692
and under Grant No. 
502254252, 
and
by the Paderborn Center for Parallel 
Computing (PC$^2$) within the project 
HPC-PRF-UBI2.
\end{acknowledgments}

\appendix
\section{Rigorous justification of the approximation (\ref{34})}
\label{app1}

As usual, the Hilbert space of the considered model 
is denoted as $\hr$, and the norm $\norm{\phi}$ of any vector $|\phi\rangle\in\hr$ 
is defined as $\langle\phi |\phi\rangle^{1/2}$.
Furthermore, for any linear
(but not necessarily Hermitian) operator $B:\hr\to\hr$,
the operator norm is defined as 
\begin{eqnarray}
\opnorm{B}:=\sup_{\norm{\phi}=1}\norm{B|\phi\rangle }
\ .
\label{a1}
\end{eqnarray}
If $B$ is Hermitian, this is equal to the maximum (supremum)
of all its eigenvalues in modulus.
Some well-established relations
for arbitrary linear (but not necessarily Hermitian)
operators $B$ and $C$ are
\begin{eqnarray}
\opnorm{B C}
& \leq &
\opnorm{B}\opnorm{C} 
\label{a2}
\\
\opnorm{B+C}
& \leq &
\opnorm{B}+\opnorm{C}
\label{a3}
\\
\opnorm{B^\dagger}
& = &
\opnorm{B}
\, ,
\label{a4}
\\
\opnorm{B^\dagger B}
& = &
\opnorm{B}^2
\, .
\label{a5}
\end{eqnarray}
For example, $\psi(\lambda)$ in (\ref{20}) is such 
a linear but generally not Hermitian operator.

Our first main goal ist to derive bounds for the quantity 
$R_t$ from Eq. (\ref{32}),
which we rewrite as
\begin{eqnarray}
R_t
& = &
\!\int_0^\beta\! dx \!\int_0^x \! dy \, r_t(x,y)
\ ,
\label{a6}
\\[0.1cm]
r_t(x,y)
& := &
\tr\left\{\rho V_x V_y \psi(y) A(t)\right\}
\ .
\label{a7}
\end{eqnarray}
Considering $\tr\{\rho B^\dagger C\}$ as a scalar product 
of
two arbitrary linear (but not necessarily Hermitian) operators $B$ 
and $C$, 
the Cauchy-Schwarz inequality takes the form 
\begin{eqnarray}
|\tr\{\rho \, B^\dagger C\}|^2
\leq 
\tr\{\rho \, B^\dagger B\} \tr\{\rho \, C^\dagger C\}
\ . \ \ 
\label{a8}
\end{eqnarray}
Moreover, for an arbitrary Hermitian, non-negative operator $D$ one readily verifies by
evaluating the trace by means of the eigenbasis of $D$ that
$|\tr\{\rho D\}|\leq \opnorm{D}\tr\{\rho\}=\opnorm{D}$.
Since $B^\dagger B$ and $C^\dagger C$ in (\ref{a8}) are Hermitian and non-negative,
it follows with (\ref{a5}) and (\ref{a8}) that
\begin{eqnarray}
|\tr\{\rho B^\dagger C\}|^2
\leq 
\opnorm{B}^2 \opnorm{C}^2
\, . \ \ \ \ \ \ 
\label{a9}
\end{eqnarray}
By exploiting this result and (\ref{a2}), we can infer from (\ref{a7}) that
\begin{eqnarray}
|r_t(x,y)|
& \leq &
\opnorm{V_y}\opnorm{V_x}\opnorm{\psi(x)} 
\opnorm{A}
\ .
\label{a10}
\end{eqnarray}
Without any significant loss of generality, we henceforth
restrict ourselves to non-negative values of $\beta$
(negative $\beta$ can be readily accounted for by changing the sign of $H$, $H_{\! g}$, and $V$ in (\ref{11})).
Eq. (\ref{a6}) together with (\ref{a10}) then yields
\begin{eqnarray}
|R_t|
& \leq &
\!\int\limits_0^\beta\! dx \!\int\limits_0^x \! dy \, |r_t(x,y)|
\leq
\frac{\beta^2}{2}\, M_V^2 M_\psi \
\opnorm{A}
\, , \ \ \ 
\label{a11}
\\
M_V
& := &
\max_{y\in[0,\beta]}\opnorm{V_y}
\, ,
\label{a12}
\\
M_\psi
& := &
\max_{y\in[0,\beta]}\opnorm{\psi(y)}
\, .
\label{a13}
\end{eqnarray}

Exploiting (\ref{24}) and (\ref{a2}), we can conclude that
\begin{eqnarray}
\opnorm{\psi'(y)}\leq g \, 
\opnorm{V_y}\opnorm{\psi(y)}
\, .
\label{a14}
\end{eqnarray}
{To deduce an upper bound for $\opnorm{\psi(y)}$ from this relation,
we proceed similarly as in \cite{vor22}:}
Observing that $|\opnorm{B}-\opnorm{C}|\leq \opnorm{B-C}$
for arbitrary $B$ and $C$ (see (\ref{a3})),
and choosing $B=\psi(y+dy)$ and $C=\psi(y)$,
it readily follows that
\begin{eqnarray}
\left| \frac{d\opnorm{\psi(y)}}{d y}\right|\leq\opnorm{\psi'(y)}
\ .
\label{a15}
\end{eqnarray}
With (\ref{a14}) and (\ref{a12}) this implies for all 
$y\in[0,\beta]$ that
\begin{eqnarray}
\left| \frac{d\opnorm{\psi(y)}}{d y}\right|
& \leq & 
g\, \opnorm{V_y} \opnorm{\psi(y)} \leq
g\, M_V \opnorm{\psi(y)}
\, . \ \ \ \ \ \ \ 
\label{a16}
\end{eqnarray}
Focusing on $y\in[0,\beta]$ and observing that the real valued function $\opnorm{\psi(y)}$ 
of $y$ is non-negative and that its growth is upper bounded by the right hand side of (\ref{a16}), 
it must be upper bounded by a function $f(y)$ which satisfies $f'(y)=M_V f(y)$ and $f(0)=\opnorm{\psi(0)}$.
Upon integrating this equation and exploiting that $\psi(0)=1$ according to (\ref{20}),
we thus obtain
\begin{eqnarray}
\opnorm{\psi(y)}\leq e^{ygM_V}
\ .
\label{a17}
\end{eqnarray}
With (\ref{a13}) this yields
\begin{eqnarray}
M_\psi \leq e^{g \beta M_V}
\ .
\label{a18}
\end{eqnarray}

In view of (\ref{a11}), (\ref{a12}), (\ref{a18}), the main remaining task is to
upper bound the operator norm of $V_y:=e^{yH}Ve^{-yH}$ (see Eq. (\ref{25})).
Such bounds have been previously obtained for a considerable
variety of Hamiltonians $H$ and perturbations $V$, 
see for instance Refs. \cite{aba15,ara16,oli18,avd20} 
and further references therein.
Focusing on Ref. \cite{avd20}, Eq. (4) therein can be rewritten 
in our present notation as
\begin{eqnarray}
\opnorm{V_y}=\opnorm{e^{yH}Ve^{-yH}}\leq \opnorm{V} f(y)
\ .
\label{a19}
\end{eqnarray}
Concerning $f(y)$, let us first consider
lattice systems in one dimension with 
nearest-neighbor interactions and 
Hamiltonians of the form 
\begin{eqnarray}
H=\sum_{l=1}^L h_l
\ ,
\label{a20}
\end{eqnarray}
see Eq. (5) in Ref. \cite{avd20} and our present Eq. (\ref{46})
for an explicit example.
Similarly as below Eq. (5) and in Eq. (14) of Ref. \cite{avd20},
we employ the definitions
\begin{eqnarray}
{\gamma} & := & \max_l\opnorm{h_l}
\ .
\label{a21}
\\
q_y & := & e^{2|y| {\gamma}} -1
\ .
\label{a22}
\end{eqnarray}
The first main result in Ref. \cite{avd20} then takes the form 
(see Eq. (19) therein)
\begin{eqnarray}
f(y)=e^{2q_{y}}
\ .
\label{a23}
\end{eqnarray}
Analogously, Eq. (39) in \cite{avd20} implies the following bound for
lattice systems in arbitrary dimensions:
\begin{eqnarray}
f(y) \leq  \frac{9}{1-2{\gamma} |y| z}
\ ,
\label{a24}
\end{eqnarray}
where $z$ is the coordination number of the considered lattice,
which means that each vertex is attached/adjacent to at most z bonds
(see above Eq. (38) and beginning of Appendix B in \cite{avd20}).
It thus seems that this result, in particular, is not restricted 
to cases with nearest-neighbor interactions.
The main new point is that only $\beta$ values with $\beta\leq1/2{\gamma}z$
are now admitted.
{In any case, the Hamiltonian $H$ must be of the form
(\ref{a20}) with local operators $h_l$
(hence $H$ itself is often denoted as a local Hamiltonian).}

More precisely speaking, the results in (\ref{a22})-(\ref{a24}) 
actually only apply if
$V$ is a local operator which acts on a single site of the lattice,
see below Eq. (5) in \cite{avd20}.
By linearity, one readily infers that the same results remain valid
if $V$ is a sum of single site operators and $\opnorm{V}$ 
in (\ref{a19}) is replaced by the corresponding sum of operator 
norms.
Moreover, for one-dimensional lattice 
Hamiltonians (\ref{a20}) with nearest-neighbor
interactions, the bound (\ref{a19})
still remains valid for operators $V$ 
which act on $N_s$ consecutive sites 
(the letter ``$s$" stands for ``support"), 
while (\ref{a23}) must now be generalized as follows \cite{dym}:
\begin{eqnarray}
f(y)=\frac{1-q_y^{N_s}}{1-q_y}\, e^{2q_y}
 \ .
\label{a25}
\end{eqnarray}
The generalization to sums of such operators is again obvious.
For lattice Hamiltonians (\ref{a20}) with more general interactions 
and in more than one dimension,  
analogous generalizations
are expected to be feasible 
as well \cite{dym}.

Next, we can conclude from (\ref{25}), (\ref{a12}), and (\ref{a19})
that 
\begin{eqnarray}
M_V
\leq
\opnorm{V} f(\beta)
\label{a26}
\end{eqnarray}
and with (\ref{a13}), (\ref{a18}) that
\begin{eqnarray}
|R_t|
& \leq &
\opnorm{A}
\, M^2 e^{g M}/2
\, , \ \ \ 
\label{a27}
\\
M
 f(\beta)\, & := & \beta\, f(\beta)\, \opnorm{V}
\ .
\label{a28}
\end{eqnarray}

Finally, we turn to the quantity $Q$ from (\ref{29}).
Obviously, this quantity is recovered by choosing $A=1$
in (\ref{32}) and thus in (\ref{a27}), yielding
\begin{eqnarray}
|Q| \leq  M^2 e^{g M}/2
\ .
\label{a29}
\end{eqnarray}

It is instructive to rewrite (\ref{a27})-(\ref{a29}) as
\begin{eqnarray}
g^2 |R_t| & \leq & \opnorm{A} \, F(\epsilon)
\ ,
\label{a30}
\\
g^2 |Q| & \leq & F(\epsilon)
\ ,
\label{a31}
\\
F(x) & := & x^2 e^x/2
\ ,
\label{a32}
\\
\epsilon & := & g\beta f(\beta) \opnorm{V}
\ .
\label{a33}
\end{eqnarray}
We also recall that $g\beta \opnorm{V}$ quantifies the
perturbation strength in (\ref{11}) in units of the thermal energy,
and that $f(\beta)$ is of order unity 
at least for small-to-moderate values of $\beta$
if we work in natural energy units with $J=1$ in (\ref{a21}).

The main conclusion is that the right hand side of (\ref{a30}) 
is independent of $t$.
Moreover, $g^2 R_t$ in (\ref{a30}) and $g^2 Q$ in (\ref{a31})
scale asymptotically (at least) quadratically 
with $g$, $\beta$, and $V$.

Altogether, 
(\ref{a30})-(\ref{a33}) thus amounts
to a rigorous justification of the
approximation (\ref{34}) in the main text
(see also discussion at the beginning 
of Sec. \ref{s3}).

\section{Derivation of the generalization from Sec. \ref{s6}}
\label{app2}

As in Sec. \ref{s6}, we take for granted that the Hamiltonian 
$H$ and the perturbation $V$ in (\ref{11}) obey the relations
(\ref{55}), where $P$ is a projector onto some subspace
$\tilde\hr$.

By exploiting (\ref{55}) and utilizing the common 
eigenbasis of $H$ and $P$, one readily verifies that 
\begin{eqnarray}
Pe^{zH}=e^{zH}P=Pe^{zH}P
\label{b1}
\end{eqnarray}
for any $z\in\CC$, and similarly for $H_g$ from (\ref{11}),
\begin{eqnarray}
Pe^{zH_g}=e^{zH_g}P=Pe^{zH_g}P
\ .
\label{b2}
\end{eqnarray}
Defining the projected/reduced Hamiltonians as
\begin{eqnarray}
\tilde H := PHP\,,\ \tilde H_g := PH_gP
\ ,
\label{b3}
\end{eqnarray}
one furthermore can infer that
\begin{eqnarray}
Pe^{zH}=Pe^{z\tilde H}=e^{z\tilde H}P=Pe^{z\tilde H}P
\ ,
\label{b4}
\end{eqnarray}
and likewise for $\tilde H_g$.
On the other hand, relations such as, for instance, 
$Pe^{zH}P=e^{zPHP}$
are generally {\em not} valid\,!
(This is particularly obvious for $z=0$.)

Similarly as in (\ref{b3}), we define the following projected/reduced 
counterparts of the original quantities introduced in Sec. \ref{s1}:
\begin{eqnarray}
\tilde A & := &PAP\,,\ \tilde V:=PVP\,,
\label{b5}
\\
\tilde \rho & := & Pe^{-\beta \tilde H}P/\, \tr\{Pe^{-\beta \tilde H}P\}
\ ,
\label{b6}
\\
\tilde A_{th} & := & \tr\{\tilde \rho \tilde A\}
\ ,
\label{b7}
\\
\tilde A(t)
& := &
P e^{i\tilde H t/\hbar} \tilde A e^{-i\tilde Ht/\hbar} P
\ ,
\label{b8}
\\
\tilde C_{\! V\!\!A}(t) & := &\tr\{\tilde \rho\;\! \tilde V\! \tilde A(t)\} - \tilde V_{th} \tilde A_{th}
\ ,
\label{b9}
\\
\tilde \rho_0 & := & P e^{-\beta \tilde H_{\! g}}P/\,\tr\{Pe^{-\beta \tilde H_{\! g}}P\}
\ ,
\label{b10}
\\
\widetilde{\langle A\rangle}_t & := &\tr\{\tilde \rho_0 \tilde A(t)\} 
\ .
\label{b11}
\end{eqnarray}
Essentially, everything is thus reduced/projected  
onto the subspace $\tilde\hr$. 

The key point of this Appendix consist in the claim
that our main result (\ref{16}) 
still remains valid
when working
from the outset within the subspace $\tilde\hr$,
and assumes the form
\begin{eqnarray}
\widetilde{\langle A\rangle}_t - \tilde A_{th}
 & = &  
g \beta \sum_{k=0}^\infty  \frac{(i\tau_B)^k}{(k+1)!}\,  \tilde C^{(k)}_{V\! A}(t)
\ .
\label{b12}
\end{eqnarray}

To substantiate this claim, we simply must replace all the 
quantities in Sec. \ref{s1} and Appendix \ref{app1} by 
their above defined reduced counterparts.
The only major problem is that the reduced Hamiltonian
$\tilde H$ and perturbation $\tilde V$ generally no longer exhibit
the specific properties of their original counterparts $H$ and $V$
which are required for instance around (\ref{a20}), (\ref{a24}), 
and (\ref{a25}).
This problem can be solved by noting that the actual
remaining task in our present case is, analogously
to the sentence below (\ref{a18}), to upper bound the
operator norm of
\begin{eqnarray}
\tilde V_y:=Pe^{y\tilde H}\tilde Ve^{-y\tilde H}P
\ .
\label{b13}
\end{eqnarray}
By exploiting (\ref{b4}) and (\ref{b5}) it follows that
\begin{eqnarray}
\tilde V_y=Pe^{y  H} V e^{-y  H}P
\ ,
\label{b14}
\end{eqnarray}
and with (\ref{25}), (\ref{a2}) that
\begin{eqnarray}
\opnorm{\tilde V_y}\leq\opnorm{V_y}\opnorm{P}^2
\ .
\label{b15}
\end{eqnarray}
Recalling the definition (\ref{a1}) and the fact that $P$ is a projector,
it follows that $\opnorm{P}=1$ and hence
\begin{eqnarray}
\opnorm{\tilde V_y}\leq\opnorm{V_y}
\ .
\label{b16}
\end{eqnarray}
Accordingly, we can again exploit (\ref{a19}) and all the results for
$f(y)$ as detailed below (\ref{a19}).
Altogether, this completes the justification 
({\ref{a12}).

Finally, we note that some of the projectors $P$
and tilde symbols in (\ref{b6}), (\ref{b8}), (\ref{b10})
are actually superfluous according to (\ref{b1})-(\ref{b4}).
In particular, one readily confirms that
\begin{eqnarray}
\tilde \rho 
& = & Pe^{-\beta H}/\, \tr\{Pe^{-\beta H}\}
\ ,
\label{b17}
\\
\tilde A(t)
& = &
P e^{iH t/\hbar} A e^{-i Ht/\hbar} P= PA(t)P
\ ,
\label{b18}
\\
\tilde \rho_0 
& = &
P e^{-\beta H_{\! g}}/\,\tr\{Pe^{-\beta H_{\! g}}\}
\ ,
\label{b19}
\end{eqnarray}
where $A(t)$ is defined in (\ref{6}).
Moreover, (\ref{b7}), (\ref{b9}), (\ref{b11})
can be rewritten as
\begin{eqnarray}
\tilde A_{th} & := & \tr\{\tilde \rho A\}
\ .
\label{b20}
\\
\tilde C_{\! V\!\!A}(t)
& = &
\tr\{\tilde \rho\,  V\! A(t)\} - \tilde V_{th} \tilde A_{th}
\ .
\label{b21}
\\
\widetilde{\langle A\rangle}_t 
& = &
\tr\{\tilde \rho_0  A(t)\} 
\ .
\label{b22}
\end{eqnarray}
It follows that (\ref{b12}) assumes the same form as
(\ref{16}) if the canonical ensembles 
$\rho$ and $\rho_g$ from (\ref{1}) and (\ref{9}) 
are replaced by their counterparts 
$\tilde \rho$ and $\tilde \rho_g$ from (\ref{b17}) and (\ref{b19}). 
Analogous modifications of (\ref{7}), (\ref{8}), 
(\ref{18}), (\ref{19}), as well as of Onsager's hypothesis
(\ref{15}) are obvious and therefore omitted.
Upon dropping the tilde symbols in (\ref{b17}) and (\ref{b19}),
our verification of the statement above Eq. (\ref{57}) is 
thus completed.



\begin{thebibliography}{99}
\bibitem{f0}
We tacitly restrict ourselves to model systems which
exhibit a well-defined thermodynamic limit.
\bibitem{f1}
Rigorously speaking, one can readily upper and lower bound
the quantities $\langle A\rangle_t$ and $A_{th}$
in terms of the operator norm of $A$, 
which is independent of the system size.
The remaining possibility that these quantities
nevertheless do not converge in the thermodynamic 
limit appears quite unlikely.
\bibitem{f2}
An upper 
bound for $|C_{\! V\!\!A}(t)|$ in (\ref{5}),
which is moreover independent of the system 
size, can be deduced at least for $t=0$ 
from the so-called cluster decomposition property.
The latter is known to apply to the canonical ensembles 
of a large class of system Hamiltonians $H$ in (\ref{1}).
For finite $t$, similar conclusions may possibly be inferred
by invoking Lieb-Robinson bounds.
In view of such system size-independent bounds 
for $|C_{\! V\!\!A}(t)|$ it 
seems reasonable to expect that the quantity 
$C_{\! V\!\!A}(t)$ itself asymptotically approaches
a well-defined (finite) thermodynamic limit.
\bibitem{f3}
In many cases, one important such characteristic 
time scale is related to the relaxation behavior 
(often an exponential decay) for sufficiently late times.
\end{thebibliography}
\end{document}